\documentclass[letterpaper,twocolumn,10pt]{article}
\usepackage{usenix,epsfig,endnotes}
\usepackage{url}
\usepackage{hyperref}
\usepackage{booktabs}
\usepackage{amsmath}
\usepackage{amssymb}
\usepackage{dblfloatfix}
\usepackage{multirow}

\newcommand{\etal}{{\textit{et al.}}}
\newcommand{\eg}{{\textit{e.g.}}}
\newcommand{\ie}{{\textit{i.e.}}}

\usepackage{ifthen}

\newboolean{print_todo}
\setboolean{print_todo}{true}

\ifthenelse{\boolean{print_todo}} {
  \newcommand{\todo}[1]
  {\marginpar{\baselineskip0ex\rule{1.1cm}{0.5pt}\\[0ex]{\tiny\textsf{#1}}}}
}{
  \newcommand{\todo}[1]
  {}
}

\begin{document}

\date{}

\title{\Large \bf PIN Skimming: Exploiting the Ambient-Light Sensor in Mobile Devices}

\author{
{\rm Raphael Spreitzer}\thanks{This work has been supported by the Austrian Research Promotion Agency (FFG) and the Styrian Business Promotion Agency (SFG) under grant number 836628 (SeCoS). We would also like to thank Peter Teufl and Horst Possegger for helpful discussions about machine learning. Last but not least, we would like to thank the volunteers for their participation in our experiments.}\\
IAIK, Graz University of Technology, Austria \\
raphael.spreitzer@iaik.tugraz.at
} 

\maketitle


\subsection*{Abstract}
In this paper, we propose a new type of side channel which is based on the ambient-light sensor employed in today's mobile devices. The pervasive usage of mobile devices, \ie, smartphones and tablet computers and their vast amount of sensors represent a plethora of side channels posing a serious threat to the user's privacy and security. While recent advances in this area of research focused on the employed motion sensors and the camera as well as the sound, we investigate a less obvious source of information leakage, namely the ambient light. We successfully demonstrate that minor tilts and turns of mobile devices cause variations of the ambient-light sensor information. Thus, we are the first to show that this sensor leaks sensitive information. Furthermore, we demonstrate that these variations leak enough information to infer a user's personal identification number (PIN) input based on a set of known PINs. Our results show that we are able to determine the correct PIN---out of a set of 50 random PINs---within the first ten guesses about 80\% of the time. In contrast, the chance of finding the right PIN by randomly guessing ten PINs would be 20\%. Since the data required to perform such an attack can be gathered without any specific permissions or privileges, the presented side channel seriously jeopardizes the security and privacy of mobile-device owners.

\section{Introduction}
Mobile devices such as smartphones and tablet computers have become a ubiquitous part of our everyday life. Powerful processors and a variety of sensors led to manifold applications being developed on these devices. Besides more general applications that allow users to browse the Internet as well as to take and view pictures, more sophisticated applications like augmented-reality applications and games also exist. However, these devices are not solely employed for entertainment applications but also for handling business applications such as banking transactions and payment services. This in turn leads to sensitive information being processed on these devices, which also attracts the attention of criminals and imposters aiming to steal money from users and to spy on specific people through malicious applications. Hence, the investigation of security and privacy compromising threats is of utmost importance. 

One potential source of security and privacy compromising threats is denoted to side channels. Side channels represent unintended information leakage during the operation of a device and potentially allow attackers to recover secret information, as demonstrated by Kocher~\cite{DBLP:conf/crypto/Kocher96} in 1996. Mobile devices include a myriad of features and sensors that allow for many different attacks. While traditional side-channel attacks against smartcards require expensive hardware, smartphones employ many different types of sensors that already represent a plethora of side channels themselves. Thus, the ``hardware'' required to perform side-channel attacks is provided for free. 

In other words, besides providing useful information, these sensors also represent a variety of threats to the user's security and privacy. Though Android employs a permission system to prevent malicious access to specific device resources, many of the employed sensors do not require any permission at all. This exacerbates the severity of these vulnerabilities since applications without any specific permissions might be able to exploit these side channels, something that has already been demonstrated impressively. For instance, Marquardt~\etal~\cite{DBLP:conf/ccs/MarquardtVCT11} developed an application intended to run on a smartphone that is capable of recovering keyboard inputs of nearby keyboards. In this case, the application exploits the vibrations recorded via the accelerometer sensor. In 2012, Aviv~\etal~\cite{DBLP:conf/acsac/AvivSBS12} demonstrated the possibility of extracting the personal identification number (PIN) and the pattern unlock mechanism by exploiting the information provided by the accelerometer sensor in smartphones. In addition, Miluzzo~\etal~\cite{DBLP:conf/mobisys/MiluzzoVBC12} have shown how to extract pressed keys from the accelerometer sensor in combination with the gyroscope sensor. In 2013, Simon and Anderson~\cite{Simon:2013:PSI:2516760.2516770} demonstrated how to infer PINs through the camera and the microphone. 
However, compared to the work of Simon and Anderson, our attack does not require any permission and it does not signal the capturing of sensor data via audio-visual feedback. For instance, like the shutter sound or the LED while taking images with the camera. 

The importance of research in this area has been emphasized by Becher~\etal~\cite{DBLP:conf/sp/BecherFHHUW11}. In order to raise awareness about such attacks and to develop effective countermeasures, these vulnerabilities must be reported and analyzed extensively. To this end, we show how the ambient light of the user's environment seriously affects the system's security and privacy. This is due to the fact that today's mobile devices are also equipped with a light sensor that provides information about the ambient light. The most common usage of this sensor is to adjust the screen brightness depending on the light intensity. However, due to slight tilts and turns during the operation of the device, the information provided by the sensor allows an attacker to infer the input provided by the user. In this paper, we investigate the sensitive information leaked through this sensor.

\paragraph{Contribution.}
The contributions of this work can be summarized as follows. First, we show that the light sensor employed in today's mobile devices actually represents a new type of side channel that leaks the user's input like the secret PIN input. Second, we observed that the light sensor of modern smartphones also captures the red, green, blue, and white (RGBW) intensities which leaks even more information and improves the accuracy of the attack. Third, we provide practical insights into the ways in which this side channel can be exploited to gather the secret PIN input of the user by employing machine-learning algorithms. Fourth, we discuss potential mitigation techniques to prevent the exploitation of sensor-based side channels. 

\paragraph{Outline.} 
The remainder of this paper is structured as follows. In Section~\ref{sec:investigation_of_light_sensor} we start with a basic investigation of the ambient-light sensor and provide an insight into the information leaked through this sensor. Section~\ref{sec:attack_scenario} outlines one potential attack scenario which takes advantage of this side channel to recover the secret PIN input provided by the user. Later on, in Section~\ref{sec:actual_exploitation_of_light_sensor}, we detail how the leaked information is actually exploited by employing machine-learning algorithms. In Section~\ref{sec:evaluation} we extensively analyze the information gathered and provide detailed evaluations regarding the applicability and reliability of this attack. We cover limitations of the presented attack in Section~\ref{sec:limitations} and we provide a brief analysis of countermeasures to prevent such attacks in Section~\ref{sec:analysis_countermeasures}. Last but not least, we discuss related work in Section~\ref{sec:related_work} and present a conclusion for this paper in Section~\ref{sec:conclusion}. 

\begin{figure}[!t]
\centering
\includegraphics[width=2.4in]{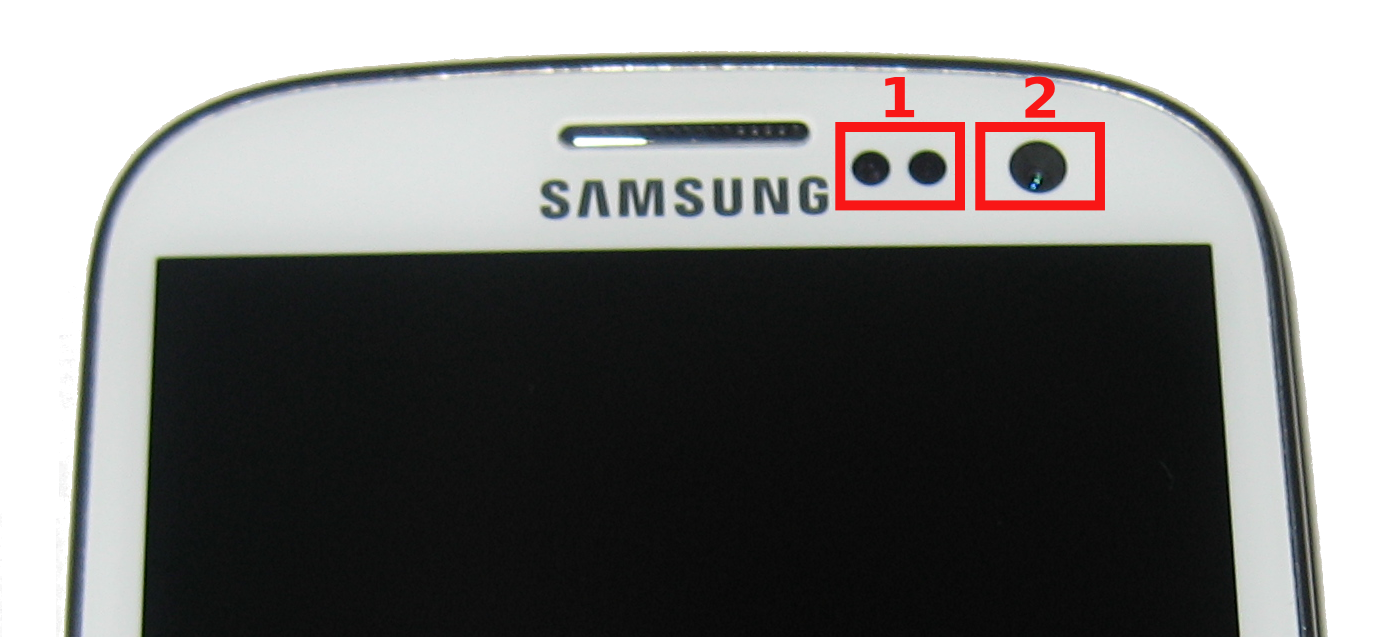}
\caption{Samsung Galaxy SIII with (1) the proximity and ambient-light sensor as well as (2) the front camera.}
\label{fig:s3_light_sensor}
\end{figure}

\section{Ambient-Light Sensor}
\label{sec:investigation_of_light_sensor}
The ambient-light sensor employed in many of today's mobile devices provides information about the intensity of the surrounding illumination. Figure~\ref{fig:s3_light_sensor} illustrates the proximity and ambient-light sensor as well as the front camera on a \emph{Samsung Galaxy SIII}, which acts as our test device. The information reported from the ambient-light sensor is given in SI \emph{lux} units, which measures the intensity of illumination of a surface. This information is used to adapt the screen brightness appropriately. For instance, outside in direct sunlight the screen brightness must be increased to remain readable, whereas in darker surroundings the screen is dimmed to reduce eye fatigue~\cite{SamsungGalaxyS4Blog}. 

\setlength{\tabcolsep}{6pt}
\begin{table}[t]
 \centering
\resizebox{3.12in}{!}{
  \begin{tabular}{lr}
 \toprule
\multicolumn{1}{c}{Rate parameter} & Sample rate \\
 \midrule
SensorManager.SENSOR\_DELAY\_FASTEST (0) & $\sim 750$\,Hz \\ 
SensorManager.SENSOR\_DELAY\_GAME (1)    & $\sim 49$\,Hz \\ 
SensorManager.SENSOR\_DELAY\_UI (2)      & $\sim 15$\,Hz \\
SensorManager.SENSOR\_DELAY\_NORMAL (3)  & $\sim 5$\,Hz \\
 \bottomrule
 \end{tabular}
}
\caption{Sampling rates on the Samsung Galaxy SIII.}
\label{tab:sampling_frequency}
\end{table}

The ambient-light sensor can be accessed via the \emph{Android Sensor API}~\cite{AndroidDevelopersSensors} that allows applications to register listeners which are notified about changes of the sensor values. The \emph{android.hardware.SensorManager.registerListener(...)} method accepts a \emph{rate} parameter which determines how fast the events should be reported. Table~\ref{tab:sampling_frequency} shows the resulting sampling frequencies for the different rate parameters according to our observations on the \emph{Samsung Galaxy SIII} smartphone running Android 4.3. We determined the number of reported values over a period of 10 seconds for each of the listed parameters and we observed that about 750 measurement samples can be gathered per second. We also observed a sensor resolution of 1 \emph{lux}, \ie, the smallest detectable change is 1 \emph{lux}. Furthermore, the ambient-light sensor can be accessed without any specific permission which allows malicious applications to access this sensor information without raising any suspicion. 

\begin{figure}[!t]
\centering
\includegraphics[width=1.5in]{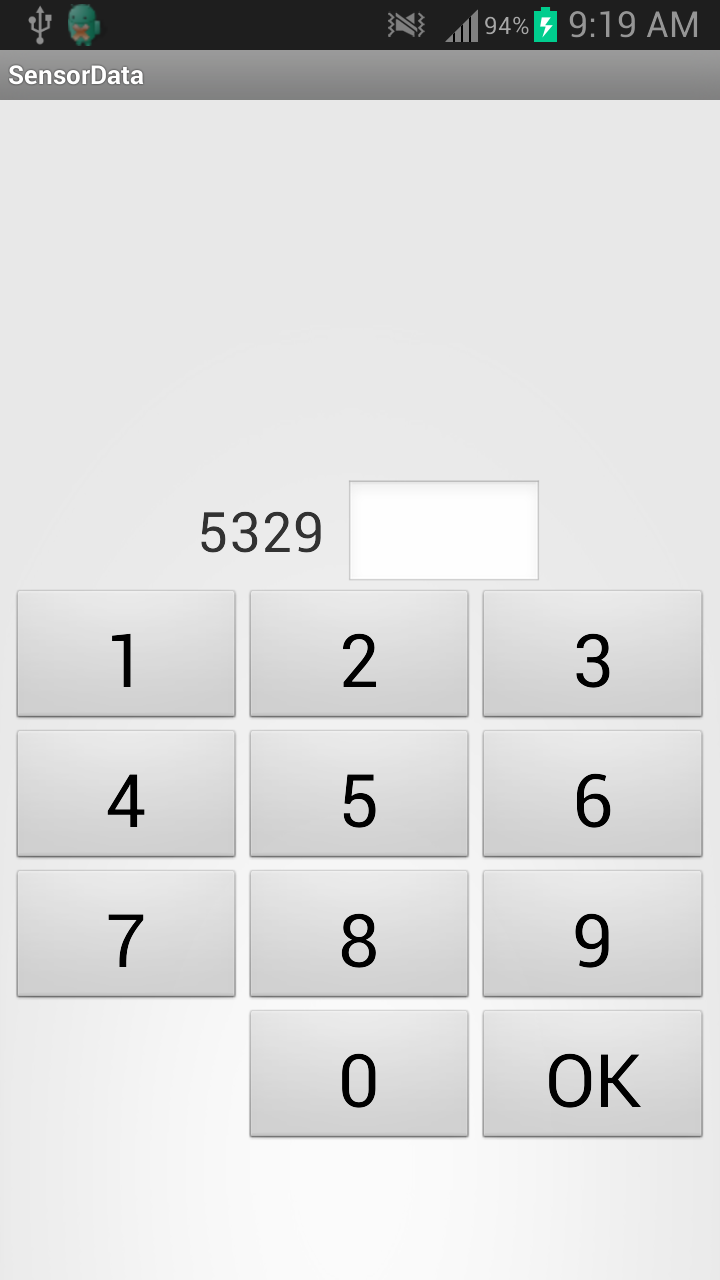}
\caption{PIN input mask to gather test samples.}
\label{fig:pin_input_mask}
\end{figure}

\begin{figure}[!t]
\centering
\includegraphics[width=2.75in]{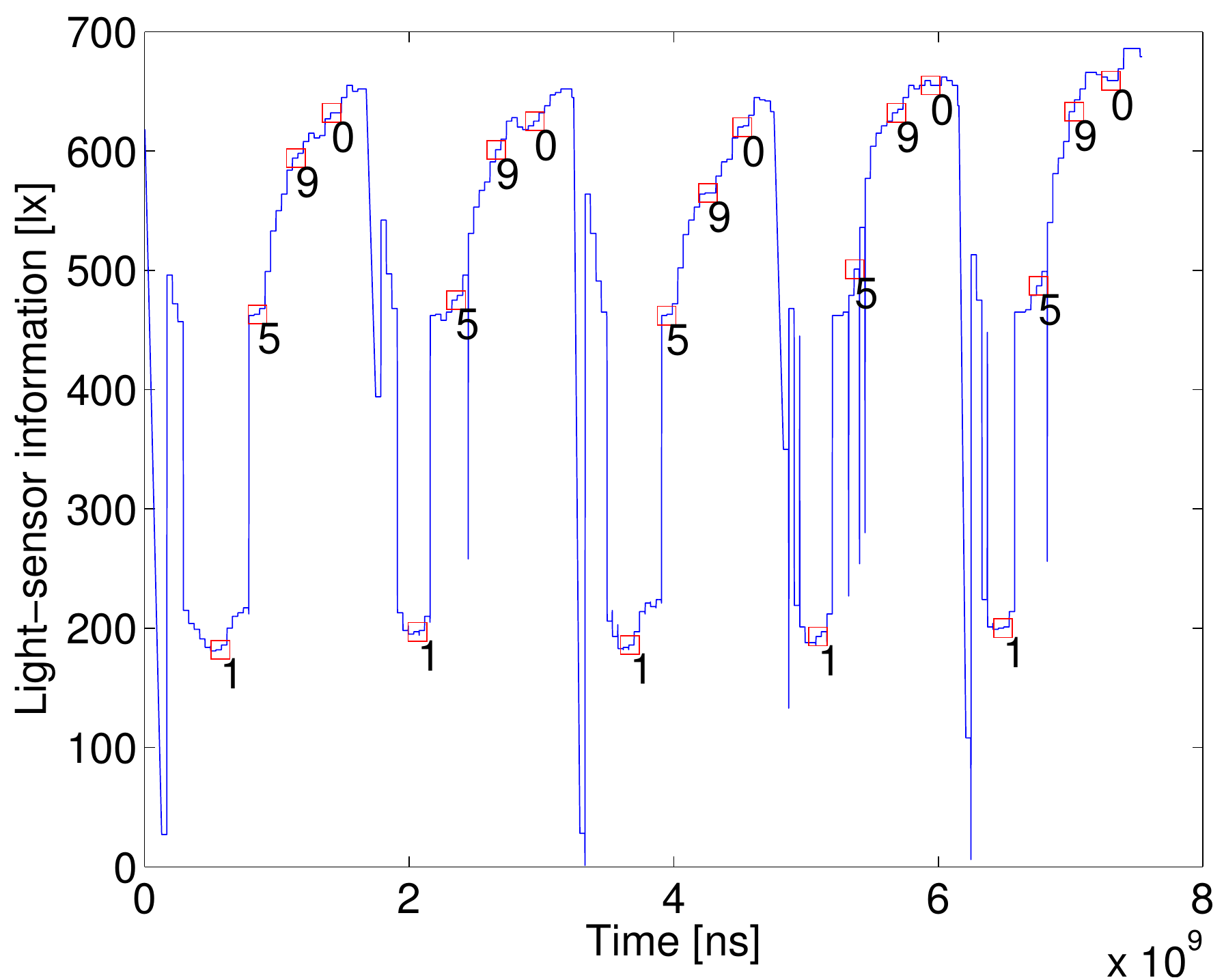}
\caption{Light-sensor information for five consecutive PIN inputs (1-5-9-0).}
\label{fig:pin_input_1590}
\end{figure}

\begin{figure}[t!]
\centering
\includegraphics[width=2.01in]{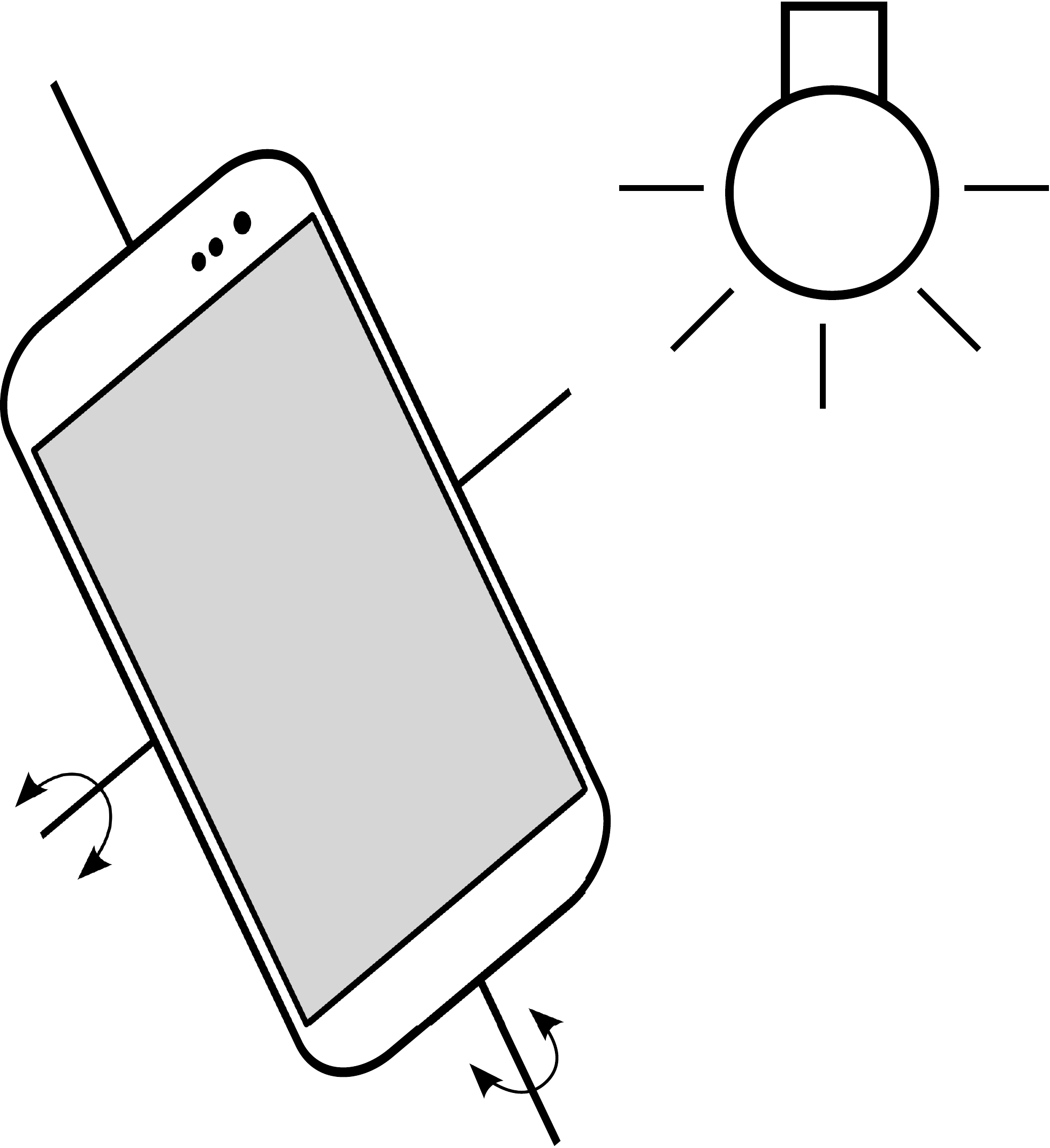}
\caption{Schematic of information leakage caused by tilts and turns of the smartphone.}
\label{fig:schematic_attack_compact}
\end{figure}

The next step in determining the applicability of the light-sensor information regarding side-channel attacks is to test the sensitivity, \ie, whether operating the smartphone actually results in changes of the information provided by the sensor. Therefore, we developed an Android application that randomly generates a 4-digit PIN and expects the user to enter this PIN on the provided PIN pad. Figure~\ref{fig:pin_input_mask} illustrates a screenshot of this application. During the input of the PIN the application collects the information provided by the ambient-light sensor as well as the corresponding timestamp of the reported value. Furthermore, we store a timestamp when a button is clicked as well as the digit of the clicked button itself. We collected this information for five consecutive PIN inputs and visualized the gathered information in Figure~\ref{fig:pin_input_1590}. For the sake of clarity we also plotted the corresponding digits of the PIN and, as can be seen in this plot, a recurring pattern for different digits of the PIN (1-5-9-0) can be observed. These differences in the light intensity during the input of the PIN occur inevitably due to slight tilts and turns while operating the smartphone. Figure~\ref{fig:schematic_attack_compact} shows a schematic illustrating the tilts and turns leading to variations of the reported sensor information. For instance, assuming the light bulb being the main source of light, then tilting the device to the left causes a decrease of the reported \emph{lux} value. Although we illustrate a point-like light source, the attack also works for environments that are uniformly lit via tube lights.

Note that the PIN input mask can be aligned on the top or on the bottom of the screen. For instance, on our test device the SIM-unlock PIN mask is aligned on the top while the standard phone pad is aligned on the bottom. However, our observations showed that the alignment does not influence the information leakage. 

Accessing the ambient-light sensor does not require special permissions and, hence, any malicious application can gather the required information without raising any suspicion. Contrary to the shutter sound or a LED that indicate an active camera~\cite{AndroidDevelopersCamera}, there is not even an audio-visual feedback that signals the user that data is being collected with the sensor. Thus, we conclude that the information provided by the ambient-light sensor can be exploited in side-channel attacks, which means that an attacker is able to determine the corresponding input based on this information.

\subsection{RGB(W) Sensor}
More recent smartphones, like the \emph{Samsung Galaxy SIII}~\cite{SamsungGalaxyS3} as well as the \emph{Samsung Galaxy S4}~\cite{SamsungGalaxyS4}, also employ an RGB(W) sensor which is capable of reporting the red, green, blue, and white (RGBW) intensities of the ambient light. Samsung refers to this sensor as an RGB sensor, but since it also reports the white intensity of the ambient light we refer to it as RGB(W) sensor. According to the official Blog of Samsung Electronics~\cite{SamsungGalaxyS4Blog}, the RGB sensor is used to optimize the screen brightness and sharpness to prevent eye fatigue. 

\begin{figure}[!t]
\centering
\includegraphics[width=2.75in]{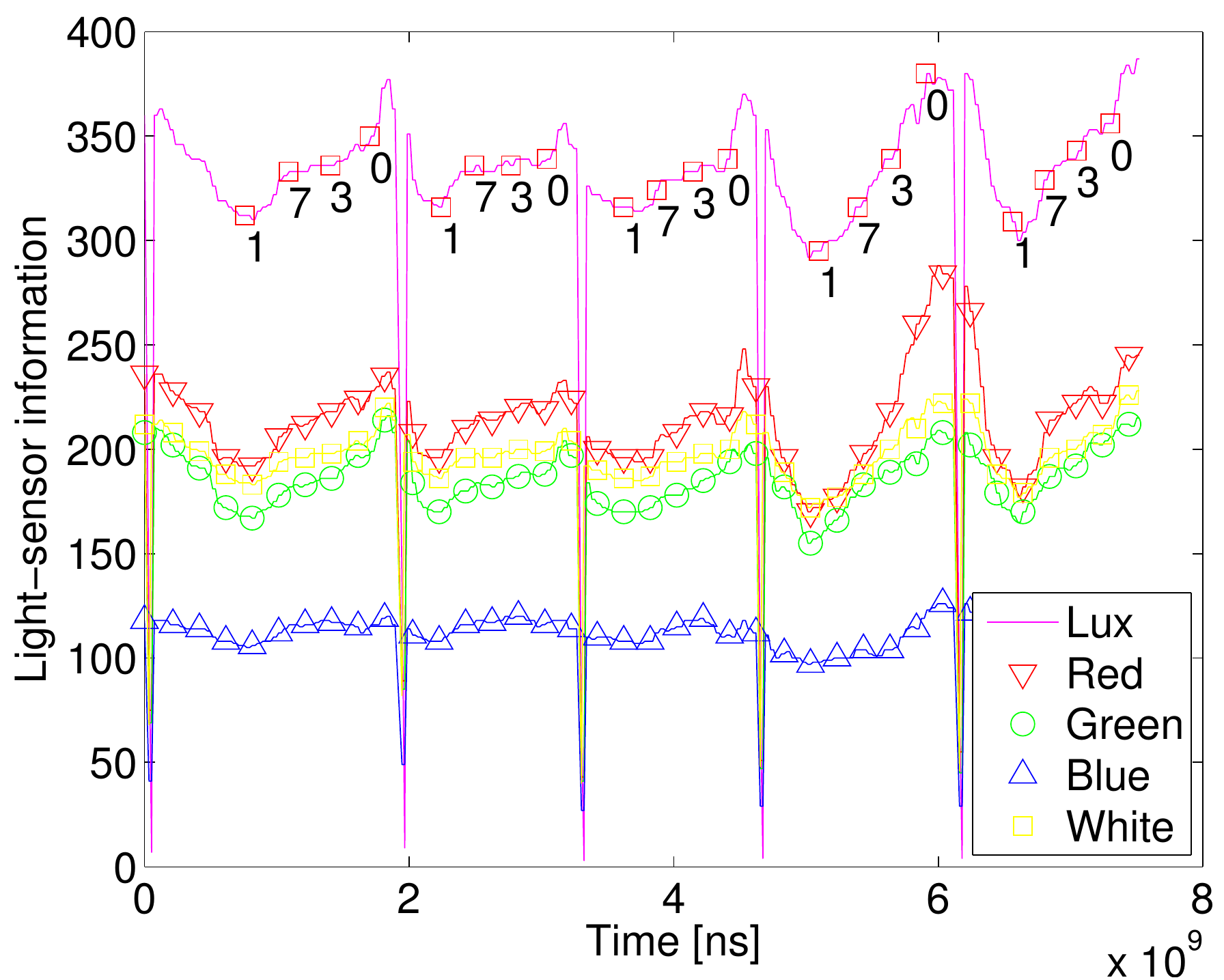}
\caption{RGB(W)-sensor information for five consecutive PIN inputs (1-7-3-0).}
\label{fig:pin_input_1730}
\end{figure}

The \emph{Android API} does not yet support RGB(W) sensors, not even the latest version (Android 4.4 \emph{KitKat}~\cite{AndroidDevelopersKitKat}). Thus, a workaround is necessary to retrieve the desired sensor values from the virtual file system directly. On the \emph{Samsung Galaxy SIII} reading \emph{/sys/devices/virtual/sensors/light\_sensor/raw\_data} yields the RGBW values of the color sensor. We verified the correctness of these values through the Service Menu~\cite{MazoGalaxyS3}---accessible by entering \verb|*#0*#| on the phone pad---which reports the same values.

Figure~\ref{fig:pin_input_1730} illustrates the data of the same experiment as mentioned above, but this time we also include the information provided by the RGB(W) sensor. Due to reasons of readability, we plotted only every 10-th value of the red, green, blue, and white intensities. We observe that all curves, \ie, red, green, blue, and white as well as the lux values, show a similar curve. However, the intensity of the blue light seems to provide a smoother curve than the other four values. Again, we plotted the event of a digit input for the sake of clarity and labeled each event appropriately.  For this specific plot we observe that the ``OK'' button leads to a rather heavy decrease of the \emph{lux} value as well as the RGBW intensities. 

Our observations show that the RGBW information provides additional information that can be exploited, \ie, additional features to be used for the machine-learning algorithm later on. Again, no specific permissions or super-user privileges are required, even though a workaround is necessary to gather the RGB(W)-sensor information. 

\section{Attack Scenario}
\label{sec:attack_scenario}
In this section, we outline one possible attack scenario. Based on this scenario, we will later on illustrate the individual steps to reproduce the actual exploitation of the light-sensor information by employing machine-learning techniques. Since the captured light-sensor information reflects the user's ambient-light conditions, a generic attack is not possible. More formally, the user needs to provide the data to be used to train the machine-learning algorithm in the same environmental setting as the actual data which is to be classified later on. Hence, we consider the following situation as a possible attack scenario. 

\subsection{Training Phase}
A hand-crafted application, \ie, an addicting game, is used to gather the light-sensor information during the user's interaction with the smartphone. The application in question tricks the user into operating the smartphone in a way that is similar to the input of multiple PINs. For example, an application similar to \emph{Math Trainer}~\cite{AndroidMathTrainer} could be used, where the user is supposed to solve mathematical puzzles and to enter numbers which can be seen as PINs. During our tests we even observed that a game where the number of correctly entered PINs in the shortest period of time could also serve as an addicting game.

We assume that users play such games in the living room while watching TV, in a waiting room while waiting for an appointment, or during a train ride. Hence, even though mobile devices are carried around all the time, there are still many opportunities to gather the required data from users operating the smartphone without walking around. Furthermore, a recent study performed by the UK's Office of Communications~\cite{OfcomStudy2013} coined the term \emph{media stacking}, which refers to the fact that about half of UK's adults conduct their smartphone or tablet computer while watching TV. Thus, our assumption seems to be reasonable and the outlined attack scenario is rather realistic. 

The game, as outlined above, might only be able to capture a limited number of ``PINs'' to be used for the classification of the unknown PIN. However, as has been shown in 2012 by Bonneau~\etal~\cite{DBLP:conf/fc/BonneauPA12} as well as Jakobsson and Liu~\cite{Jakobsson:MobileAuthentication}, people tend to choose specific PINs like dates as well as PINs that represent common four-digit words, \eg, \emph{``love''} (5683). Thus, taking research about the users' tendency to choose PINs into consideration, a set of commonly used PINs could be established and might be enough to determine the correct one. The above mentioned game can be used to learn this set of commonly used PINs. 

In case computing power for the machine-learning algorithm is required, \ie, a powerful server, the application requires an Internet connection to transmit the gathered data to this powerful intermediary. Nevertheless, we argue that convincing the user that the application requires the \emph{Internet} permission is easy. For instance, to retrieve high scores or information about new updates, new games, etc. Thus, we claim that the proposed attack can be performed without raising the user's suspicion.

\subsection{Exploitation Phase}
After gathering enough samples, the application tricks the user into restarting the device or starting the desired application, \eg, the banking application, just to capture the light-sensor information during the input of the authentication PIN. If one considers to attack the smartphone's PIN, a service can be implemented to be started on boot time. Afterwards the sensor information captured during the game play is used to deduce the unknown PIN input by means of machine learning. Note that the game, \eg, \emph{Math Trainer}, might also trick the user into buying a specific add-on, a ``new level'', or a ``new stage''. When the user performs the in-app billing via \emph{Google Wallet}, the game skims the corresponding authentication PIN. 

Now one might question whether the revealed PIN is of any value for the attacker. In fact, if the attacker later gains physical access to the mobile device, she might gain access to the mobile device by unlocking the phone, or even worse, might cause financial damage by authenticating herself to the corresponding application using the correct PIN. Furthermore, Aviv~\etal~\cite{DBLP:conf/acsac/AvivSBS12} argued that the learned PIN might be valuable in case the user reuses the PIN, for instance, as the ATM PIN. In addition, Simon and Anderson~\cite{Simon:2013:PSI:2516760.2516770} predict that the number of smartphone applications requiring an authentication PIN will increase over time. Hence, users might be tempted to reuse one PIN across different applications, which exacerbates PIN-skimming vulnerabilities. Overall, we consider the presented attack as a serious threat for today's mobile devices.

\subsection{Security Implications}
Simon and Anderson~\cite{Simon:2013:PSI:2516760.2516770} state that sensor-based side-channel attacks are capable to overcome strong separation mechanisms like \emph{BlackBerry Balance}~\cite{BlackBerryBalance} or \emph{Samsung KNOX}~\cite{SamsungKNOX}. These mechanisms try to separate the ``private'' world from the ``business'' world in order to  provide greater protection of corporate data on smartphones. However, based on the leaking sensor information, an unprivileged application running in the ``private'' world of the smartphone might gain knowledge of sensitive information entered in the ``business'' world. It is important to note that the presented attack is also capable to overcome these separation mechanisms, if unprivileged applications running in the ``private'' world are allowed to access the ambient-light sensor during the operation of the ``business'' world. Furthermore, since BlackBerry smartphones also support the execution of Android applications and the \emph{BlackBerry Runtime 10.0}~\cite{BlackBerryRuntime} also supports the ambient-light sensor, a malicious application developed for Android smartphones might also harm BlackBerry-based devices. 

\subsection{Observations and Assumptions}
The scenario outlined above is based on some observations and assumptions which are detailed within the following paragraphs. Considering a PIN-input mask as illustrated in Figure~\ref{fig:pin_input_mask}, and a user operating the smartphone with only one hand, using the thumb to enter the digits, we make the following observations. Left-handed persons tilt the smartphone slightly to the left side when entering PIN digits in the middle and right column of the key pad, \ie, 2, 3, 5, 6, 8, 9. In contrast, right-handed persons tilt the smartphone slightly to the right side when entering PIN digits in the left and middle column, \ie, 1, 2, 4, 5, 7, 8. We attribute this observation to the fact that people might possibly drop the device if operating it the other way round. Furthermore, this slightly moves the display towards the thumb, \ie, the users slightly push the display towards their thumb. These tilts of the device cause the variations in the captured light-sensor information. Similarly, we assume fewer tilts and turns of the mobile device if it is held with one hand and operated with the index finger of the other hand or a stylus pen. 

\paragraph*{Assumption 1.} We assume the user is holding the mobile device in his hands rather than laying it onto a flat surface while operating it. If we would assume the mobile device is lying on a stable surface, \ie, a table, the light-sensor information would not change during the operation of the device, unless the user's hand causes the light-sensor changes.

\paragraph*{Assumption 2.}
Furthermore, we assume that the PIN is entered on a key pad similar to the one illustrated in Figure~\ref{fig:pin_input_mask} rather than a QWERTY keyboard with a single row of numbers. Examples of applications that are ``protected'' with an authentication PIN are, for instance, AppLock~\cite{AppLock}, Evernote~\cite{Evernote}, and KeepSafe~\cite{KeepSafe}, as well as mobile banking applications, \eg, Barclays~\cite{Barclays}, and NAB~\cite{NAB}, just to name a few of them. Screenshots of these applications---provided by the corresponding developers---clearly show that the authentication PIN is entered on a key pad as illustrated in Figure~\ref{fig:pin_input_mask}. Hence, this seems to be a rather fair assumption which does not have a negative impact on the attack scenario. 

\paragraph*{Assumption 3.} 
We also assume that the user operates the mobile device in an environment where the ambient-light sensor faces a sufficiently large variance of the ambient light during the operation of the device. This is not the case in completely dark environments. However, also rather gloomy environments, \ie, a room in the late afternoon without any artificial light source, can be considered for potential attack scenarios, at least in case the \emph{lux} values vary slightly during the handling of the device.

\section{Attack Approach}
 \label{sec:actual_exploitation_of_light_sensor}
In this section, we detail the steps for the exploitation of the light-sensor information. As outlined in the scenario above, we perform a matching of sensor values captured during the input of an unknown PIN to the sensor values of known PINs. In terms of machine learning this represents a classification problem, where a so-called feature vector is mapped to a finite number of labels or categories, \ie, PINs in our case. The required steps are as follows: (1) gathering the sensor values under known PINs as well as the sensor values under unknown PINs, and (2) employing machine-learning techniques to determine the unknown PINs based on the set of known PINs.

In order to perform the classification, sensor values and the corresponding PINs are used to train the machine-learning algorithm. This data is referred to as training data. The actual data that is to be classified is known as test data. We stick to the notation of Alpaydin~\cite{Alpaydin:2010:IML:1734076} and Bishop~\cite{Bishop:2006:PRM:1162264}, \ie, bold letters denote vectors and a superscript $\text{T}$ denotes the transpose of a vector. Furthermore, uppercase bold letters denote matrices. Vectors are assumed to be column vectors by default.

\subsection{Gathering the Required Data}
The gathering of the required training data during game play can be formalized as follows. The malicious application captures a list of tuples $(t, L, R, G, B, W)$, with $t$ being the timestamp and $L$, $R$, $G$, $B$, and $W$ representing the \emph{lux} information, as well as the red, green, blue, and white intensities of the ambient light. Furthermore, we capture a list of tuples $(t_p, d)$, with $t_p$ being the timestamp of a pressed digit $d \in \{0, \dots, 9\}$ of the $p$-th PIN. In our scenario one PIN consists of four consecutive tuples in this list. The two timestamps $t$ and $t_p$ allow us to properly align the event of a pressed digit with the data of the ambient-light sensor later on. 

For each PIN we extract the sequence of sensor values within the period defined by the timestamp of the first digit and the timestamp of the last digit of the PIN. In addition, one might consider a timeframe of a few milliseconds before and after the input of the first and the last digit of one PIN, which covers additional information. The resulting matrix $\textbf{M}$ for one PIN is as follows. 
\[
 \textbf{M} = \begin{bmatrix}
  \left(t, L, R, G, B, W\right)_{1} \\
\vdots \\
\left(t, L, R, G, B, W\right)_{l} \\
 \end{bmatrix}
\]
Every column within matrix $\textbf{M}$ represents the corresponding sensor information during the input of one specific PIN, except the first one which represents the timestamp at which the information was captured. Before the gathered information can actually be exploited, we normalize the sensor values appropriately. This normalization of the data is done either by dividing each value ($L$, $R$, $G$, $B$, $W$) in one column by the norm of the corresponding column vector, or by rescaling the values via, \eg, $L_i = \left(L_i - \text{min}(L)\right)/\left(\text{max}(L) - \text{min}(L)\right)$.

The data of matrix $\textbf{M}$ is then used to derive the actual feature vectors. While Owusu~\etal~\cite{DBLP:conf/wmcsa/OwusuHDPZ12} consider a large number of features and employ special feature selectors to determine the actual set of features to be used, we keep the feature space simple and stick to a limited number of features for a first investigation. These features are outlined briefly within the following paragraphs. 
\paragraph{Lux Values Only.} The first set of feature vectors we consider are the exact \emph{lux} values at the input of each specific digit of the PIN. Therefore, we use the timestamp of each input event during one PIN input to look-up the corresponding values in matrix $\textbf{M}$. We represent these values as a vector $\textbf{x} = \left[ L_1, L_2, L_3, L_4\right]^\text{T}$, where the subscript refers to a specific digit of the PIN. 
\paragraph{Lux Values Including RGBW Values.} The second set of feature vectors we consider are the exact \emph{lux} values including the red, green, blue, and white (RGBW) intensities at the input of each digit for one specific PIN, represented as $\textbf{x} = \left[ L_1,  R_1,  G_1,  B_1,  W_1,\dots, L_4,  R_4,  G_4,  B_4,  W_4\right]^\text{T}$. In the following we refer to the feature vector comprised of these five features as \emph{LRGBW}.
\paragraph{Polynomial of Degree 3.} The third possibility we consider is fitting a polynomial of degree 3 through the second column of $\textbf{M}$. More formally, through all \emph{lux} values during one PIN input. The coefficients of this polynomial $f\left(x\right) = ax^3 + bx^2 + cx + d$ are then considered as the features of a specific PIN, \ie, $\textbf{x} = \left[a, b, c, d\right]^\text{T}$. The coefficients for the red, green, blue, and white intensities---the remaining columns of $\textbf{M}$---are obtained in the same manner and appended to the feature vector $\textbf{x}$.

After gathering the data as outlined above, any set of the above outlined feature vectors is then combined into a matrix $\textbf{F}_n$ with $n$ rows, \ie, one row for each PIN. Furthermore, a label or class vector $\textbf{c}_n = \left[(d_1, d_2, d_3, d_4)_1, \dots, (d_1, d_2, d_3, d_4)_n\right]$ of tuples corresponding to the PIN-digits can be derived.
\[ 
\textbf{F}_{n} = \begin{bmatrix}
\textbf{x}_1 \\
\vdots \\
 \textbf{x}_n
\end{bmatrix}
\text{, }\textbf{c}_n = \begin{bmatrix}
       (d_1, d_2, d_3, d_4)_1 \\
       \vdots \\
       (d_1, d_2, d_3, d_4)_n
      \end{bmatrix}
\]
The matrix $\textbf{F}_n$ as well as the label vector $\textbf{c}_n$ are then used to train the classification algorithm.

\subsection{Detecting PIN Inputs}
The above outlined approach of gathering the feature vectors can be performed in a real attack because the data for training the machine-learning algorithm can be captured during game play, in which situation all of this information is available, including the timestamp of a pressed digit. However, during the input of the unknown PIN the timestamp is not known and hence we need a mechanism to determine the PIN input on a sequence of sensor values. Therefore, the approach mentioned by Miluzzo~\etal~\cite{DBLP:conf/mobisys/MiluzzoVBC12} might be used. Their idea is to move ``windows'' of a fixed length (the length of an average PIN input time) over the sensor data and try to detect the input. Another approach by Simon and Anderson~\cite{Simon:2013:PSI:2516760.2516770} suggests to use the microphone to capture the vibrations of the haptic feedback to determine when a button on the touchscreen has been pressed. While the former approach does not require any specific permission, the latter approach requires a permission to access the microphone. However, we consider the detection of PIN inputs on a sequence of sensor values as solved and beyond the scope of this work. Thus, we gather the test data, \ie, the data to be classified, in the same way as the training data.

\subsection{Determining the Unknown PIN}
After we gathered the required light-sensor information for all the PINs ($\textbf{F}_n$ and $\textbf{c}_n$) to be used for the training phase of the machine-learning algorithm, we start the actual training phase. Therefore, we employ Matlab's Statistics Toolbox~\cite{MatlabStatisticsToolbox} with its extensive features and machine-learning algorithms. We perform the outlined attack by employing a \emph{supervised learning} algorithm, which tries to learn a function and its parameters based on labeled training data. This function is later on used to determine the label of unseen data. More formally, given a set of tuples $(\textbf{x}_i, c_i)$, with $\textbf{x}_i \in \mathbb{R}^n$ being a feature vector and $c_i$ the corresponding label of the observation, the algorithm tries to infer a function $f: X \rightarrow C$, where $X \in \mathbb{R}^n$ represents the feature vector of an observation and $C$ the inferred label.

Choosing the right classification algorithm seems to be some kind of ``mystery''. For instance, related work by Miluzzo~\etal~\cite{DBLP:conf/mobisys/MiluzzoVBC12}---who try to infer input from the accelerometer sensor and the gyroscope sensor---takes the approach of employing \emph{ensemble techniques}. This means that they train several different classification algorithms, and finally they employ a majority-voting scheme to choose the final classification result. For the sake of simplicity, we investigate three different classification algorithms and compare their results afterwards. We briefly outline the chosen classification algorithms in the following paragraphs. For further information about these algorithms we refer to~\cite{Alpaydin:2010:IML:1734076, Bishop:2006:PRM:1162264}. 

\paragraph*{Multiclass Logistic Regression.}
The classifier tries to learn parameters $\textbf{w}_k$ for every class label $k \in C$, such that a vector $\textbf{x}$ is assigned to label $k$ in case $p(k|\textbf{x}) > p(j|\textbf{x})$ for all $j \neq k$, with $p$ defined as below. 
\[
p(k|\textbf{x}) = \frac{\text{exp}\left(\textbf{w}_k^{\text{T}} \cdot \textbf{x}\right)}{\sum_i \text{exp}(\textbf{w}_i{^\text{T}} \cdot \textbf{x})}
\]

\paragraph*{Discriminant Analysis.}
The classifier tries to learn parameters $\textbf{w}_k$ for every class label $k \in C$, 
such that a vector $\textbf{x}$ is assigned to label $k$ in case $f_k(\textbf{x}) > f_j(\textbf{x})$ for all $j \neq k$, with $f$ defined as below. 
\[
f_k(\textbf{x}) = \textbf{w}_k^\text{T} \cdot \textbf{x} + w_{k0}
\] 
More formally, when talking about \emph{discriminant analysis} we refer to the \emph{linear discriminant analysis}, where classes are separated linearly. 

\paragraph*{K-Nearest Neighbor Algorithm.}
The algorithm assigns the input vector $\textbf{x}$ a label which is determined by the majority of the $K$ nearest neighbors. In our case, we used $K=5$. Usually, the Euclidean distance is used to determine the distance between two vectors.

\section{Evaluation and Results}
 \label{sec:evaluation}
We engaged a total of ten users to acquire the necessary data for the evaluation of this information leakage. Each user was asked to enter at least one set of random PINs with cardinality $k \in \{ 15, 30, 50 \}$, each PIN for a specific number of times $N \in \{3, \dots, 10\}$. If a PIN was entered incorrectly, we ignore the corresponding input and prompt the user to enter the PIN again. Unless explicitly stated, we always use the largest set of 50 PINs within the following analysis. We basically follow the approach of Aviv~\etal~\cite{DBLP:conf/acsac/AvivSBS12} who also measured the performance of their attack on a set of 50 PINs. 
In total we use the data of 29 test runs gathered by a total of 10 users.

\paragraph{Evaluation Methodology.}
Since the performance of the learned classifier could be distorted---either positively or negatively---when determined solely with the gathered test data, we apply the concept of k-fold cross validation. The purpose of this concept is to estimate the average success rate of classifying unknown data into one of the learned categories. Therefore, k-fold cross validation partitions the set of training data into \emph{k} equally-sized sets. Afterwards, $k-1$ sets of the training data are used to train the classifier and the remaining set is classified according to the learned categories to estimate the performance. This process is repeated until all $k$ possible sets have been used once as a test set. Afterwards the average performance can be estimated. Compared to the performance based solely on specific test data, the concept of cross validation provides more reliable estimations regarding the performance of a classifier~\cite{Bishop:2006:PRM:1162264}. 

\paragraph{Setup.}
We performed the experiments in rather unconstrained environments regarding the lighting conditions. This means that tests were performed in office rooms that were uniformly illuminated via tube lights, in a living room with a standard ceiling lamp, and in rooms where the only light source was a window. In the room where the only light source was a window, we even considered different daytimes, \eg, during the day and in the late afternoon. However, we asked the users not to walk around while entering the presented PINs, which is compliant with the above outlined attack scenario. Furthermore, we did not insist on a specific input method, but only that the user holds the mobile device during the operation. As already explained above, operating the mobile device while it is laying on a stable surface is unlikely to cause variations of the light-sensor information. Allowing users to choose their desired input method freely yields more generic results because the users operate the mobile device in their usual manner. 

We watched the users during the gathering of the measurement samples and observed the following input methods:
\begin{enumerate}
 \item Holding the smartphone in one hand and entering the digits with the thumb of the same hand. 
 \item Holding the smartphone in one hand and entering the digits with the thumb of the other hand.  
 \item Holding the smartphone in one hand and entering the digits with the index finger of the other hand. 
\end{enumerate}

Within the following paragraphs we analyze the gathered data with the proposed feature vectors and machine-learning algorithms in order to determine whether the secret PIN can be recovered based on a set of known PINs. 

\paragraph{Comparison of Classification Algorithms.}
As outlined above, we employ three different classification algorithms. Thus, our first intention is to determine the overall classification rate based on different classification mechanisms: (1) \emph{logistic regression}, (2) \emph{discriminant analysis}, and (3) \emph{k-nearest neighbor classification}. These classifiers are fed with the feature vectors: (a) the \emph{lux} values only (L), and (b) the \emph{lux} values including the RGBW values (LRGBW).

\begin{figure}[!t]
\centering
\includegraphics[width=2.7in]{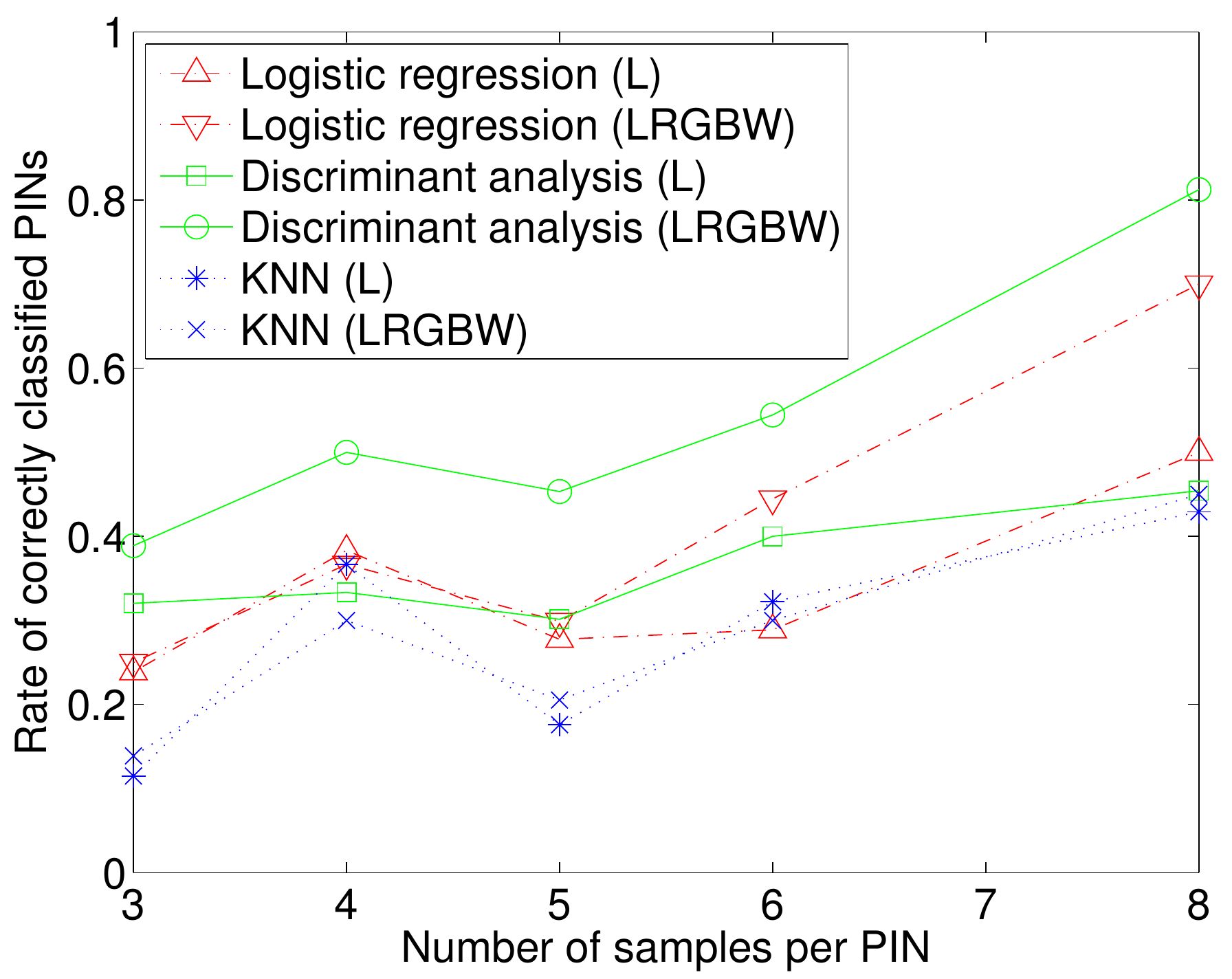}
\caption{Average rate of correctly classified PINs over multiple runs with a set of 15 PINs each.}
\label{fig:rate_of_correctly_classified_samples}
\end{figure}

We applied a 10-fold cross validation on all three classifiers and evaluated the performance of the suggested features for different numbers of samples (repetitions) per PIN. Figure~\ref{fig:rate_of_correctly_classified_samples} illustrates the average rate of correctly classified PINs out of a set of 15 known PINs for the different classifiers and the proposed feature vectors. The y-axis represents the average rate of correctly classified PINs, and the x-axis illustrates the number of gathered samples (repetitions) per PIN. Given this plot, we observe that the feature vector comprised of the LRGBW values outperforms the \emph{lux} value only feature vector most of the time for all classifiers. Thus, we conclude that the additional information leaked through the RGB(W) sensor leads to a better attack performance. We also observe that the \emph{discriminant analysis} provides better results than the other two classifiers. Furthermore, the average rate of correctly classified PINs increases with the number of samples per PIN. For instance, if we perform a \emph{linear discriminant analysis} with a training set of 15 PINs, each repeated 8 times, then we are able to classify more than 80\% of the PINs correctly. Note that the chance of correctly guessing the right PIN from a set of 15 PINs randomly is $\frac{1}{15} = 6.7\%$. 

\begin{figure}[!t]
\centering
\includegraphics[width=2.7in]{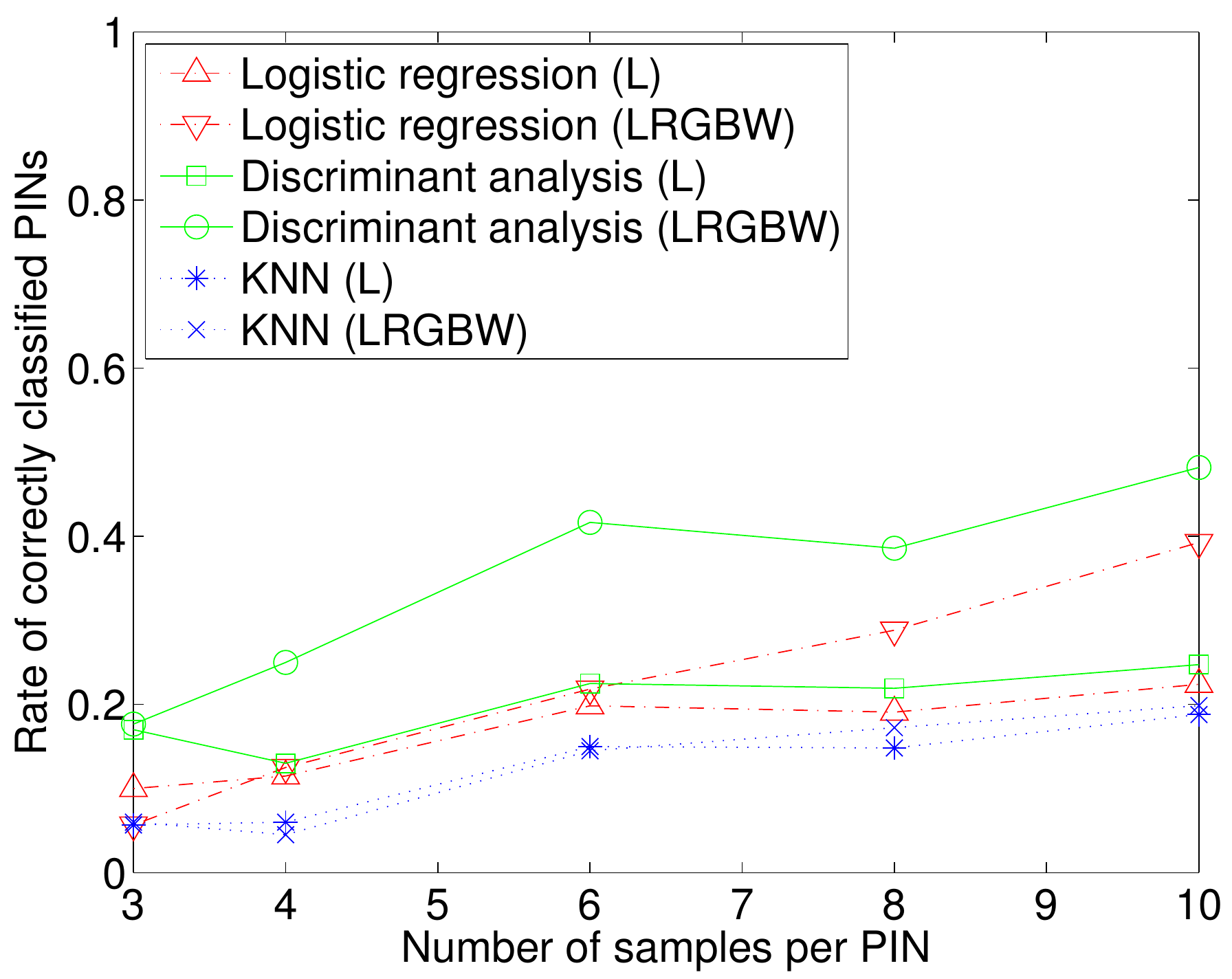}
\caption{Average rate of correctly classified PINs over multiple runs with a set of 50 PINs each.}
\label{fig:rate_of_correctly_classified_samples_50_pins}
\end{figure}

Figure~\ref{fig:rate_of_correctly_classified_samples_50_pins} illustrates the average rate of correctly classified PINs out of a set of 50 PINs. Again, the \emph{linear discriminant analysis} outperforms the other two classifiers and the additional information from the RGB(W) sensor increases the performance compared to the \emph{lux} value only. At first glance, an average rate of correctly classified PINs of 40 to 50\% seems to be quite moderate. However, the chance of correctly guessing the right PIN from a set of 50 PINs is $\frac{1}{50} = 2\%$. Thus, our attack outperforms random guessing by a factor of 20 to 25. 

\begin{figure}[!t]
\centering
\includegraphics[width=2.7in]{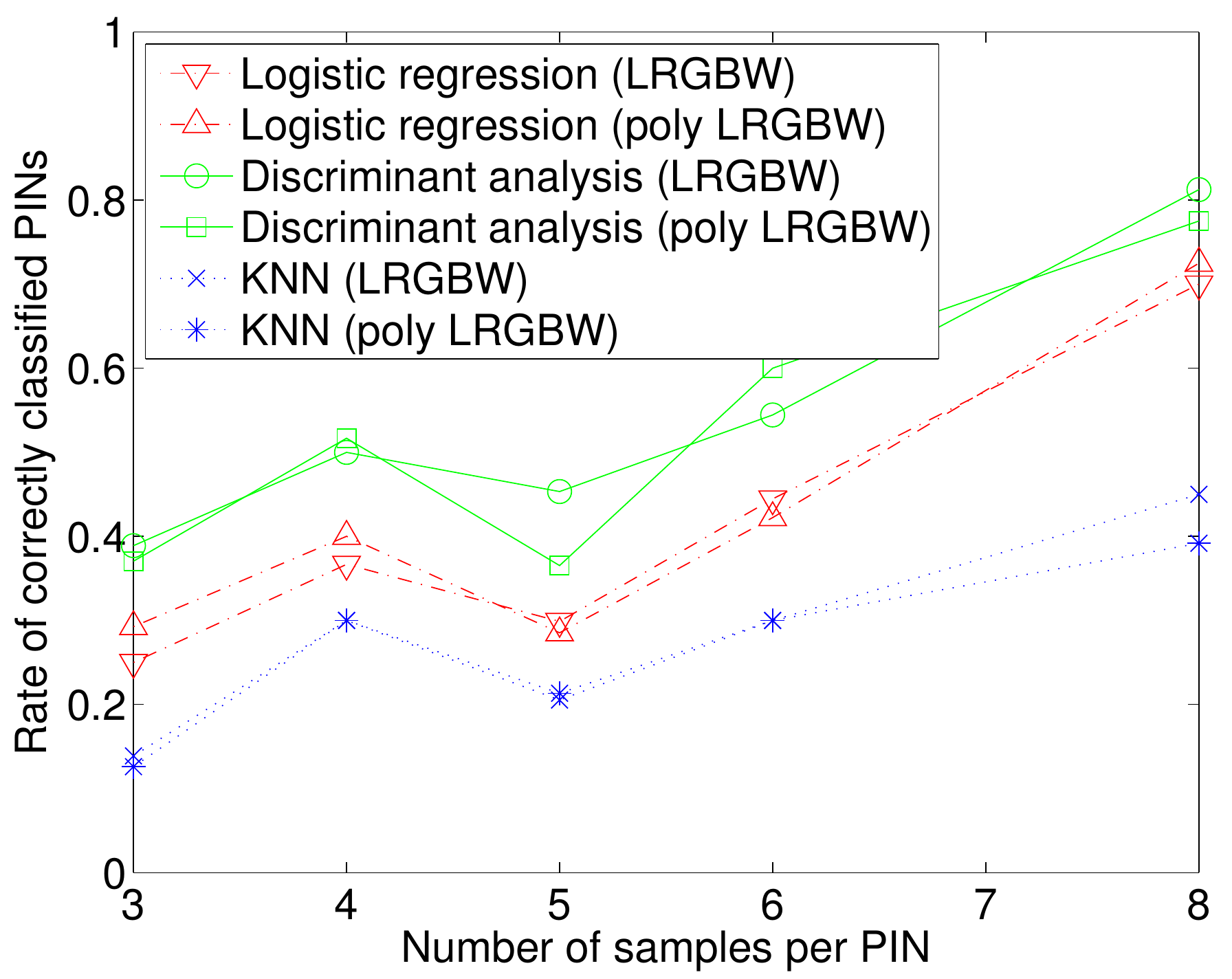}
\caption{Average rate of correctly classified PINs over multiple runs with a set of 15 PINs each.}
\label{fig:comparison_of_feature_vectors}
\end{figure}

\paragraph{Comparison of Feature Vectors.}
Since we introduced a feature vector which is comprised of a degree 3 polynomial fitted through every value (\eg, lux and RGBW values) during one PIN input, we need to compare the performance of the plain features and the approximated features. Figure~\ref{fig:comparison_of_feature_vectors} illustrates the respective performances for these feature vectors based on a 10-fold cross validation over multiple runs with a set of 15 PINs. We observe that both feature vectors yield a similar performance for the different classification algorithms, with the \emph{discriminant analysis} performing best. Based on this observation we only focus on the \emph{discriminant analysis} within the following investigations. 

\begin{figure}[!t]
\centering
\includegraphics[width=2.7in]{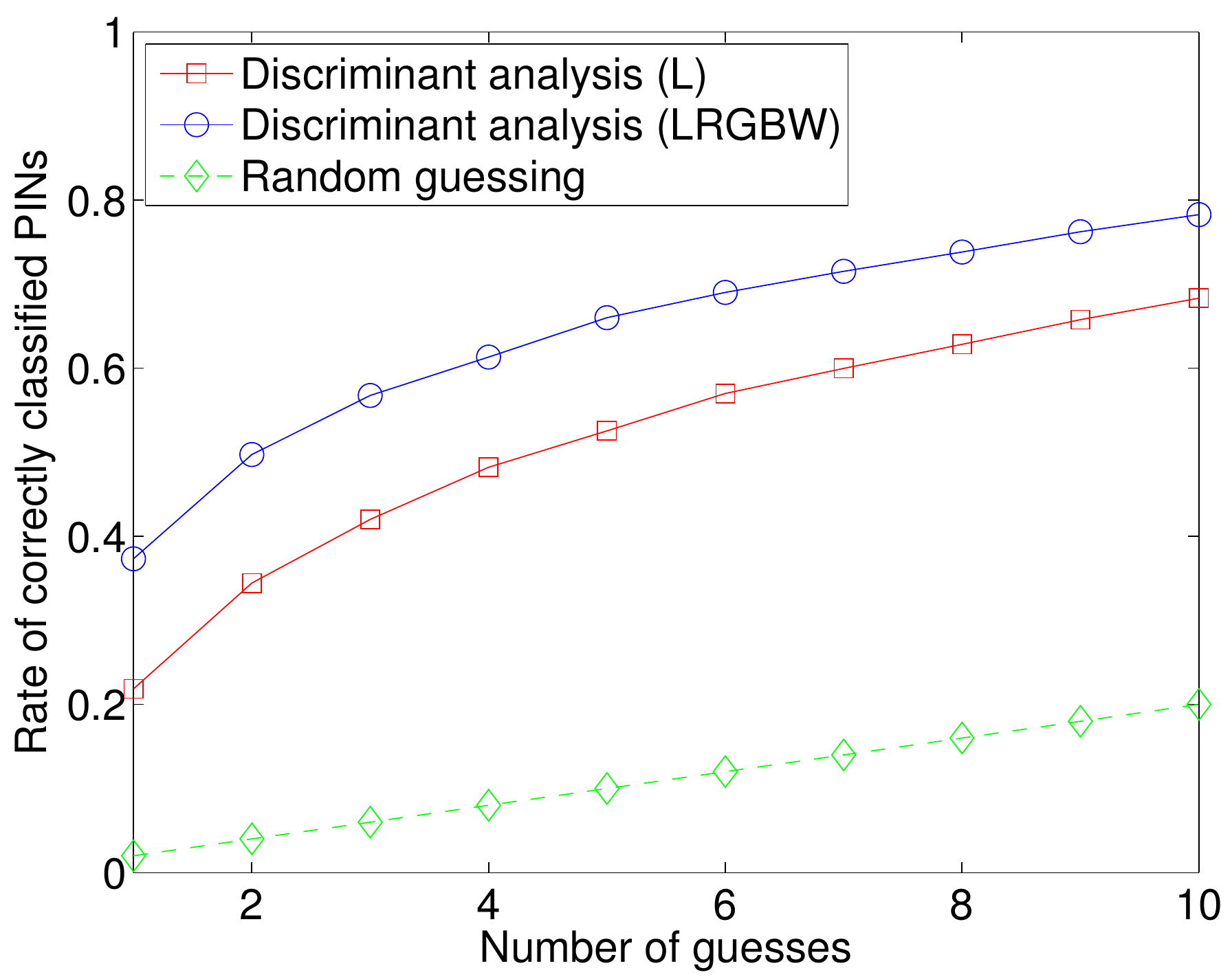}
\caption{Average rate of correctly classified PINs for multiple guesses.}
\label{fig:classify_correctly_guessing_50_pins}
\end{figure}

\paragraph{Guessing PINs According to their Probability.}
An interesting approach is to consider the fact that an adversary is able to enter PINs for a specific number of times, \ie, to guess possible PINs according to their probability for being the correct one. In this case the probability of finding the correct PIN increases with every tested PIN. 
Thus, we sort potential PINs according to their probability for being the correct one and illustrate the rate of correctly classified PINs after a specific number of guesses. 

Figure~\ref{fig:classify_correctly_guessing_50_pins} shows the average rate of correctly guessing a PIN out of a set of 50 random PINs for a specific number of guesses. The additional information provided by the RGB(W) sensor yields better results and seems to increase the success rate by about 10 percentage points compared to the \emph{lux} value only feature vector. We also illustrate the success rate if one were trying to guess PINs randomly, which clearly shows the advantage of our attack compared to random guessing. 

When comparing our results to the results of Aviv~\etal~\cite{DBLP:conf/acsac/AvivSBS12}, we observe that the ambient-light sensor provides results at least as good as those achieved by the accelerometer sensor. Comparing our results to the work of Simon and Anderson~\cite{Simon:2013:PSI:2516760.2516770}, we observe that the ambient-light sensor provides even better results than the approach of exploiting the camera to infer PINs. For instance, Simon and Anderson claim to infer more than 30\% of the PINs after two guesses and more than 50\% of the PINs after five guesses. In contrast, the ambient-light sensor allows us to infer about 50\% of the PINs after two guesses and about 65\% of the PINs after five guesses. 

The presented results indicate that guessing PINs according to their probability provides an effective means of finding the correct one. On average we are able to determine the correct PIN with a probability of 80\% when considering the ten most probable PINs. In contrast, guessing PINs randomly from a set of 50 PINs would result in a success rate of 20\% after ten guesses.

\begin{figure}[!t]
\centering
\includegraphics[width=2.7in]{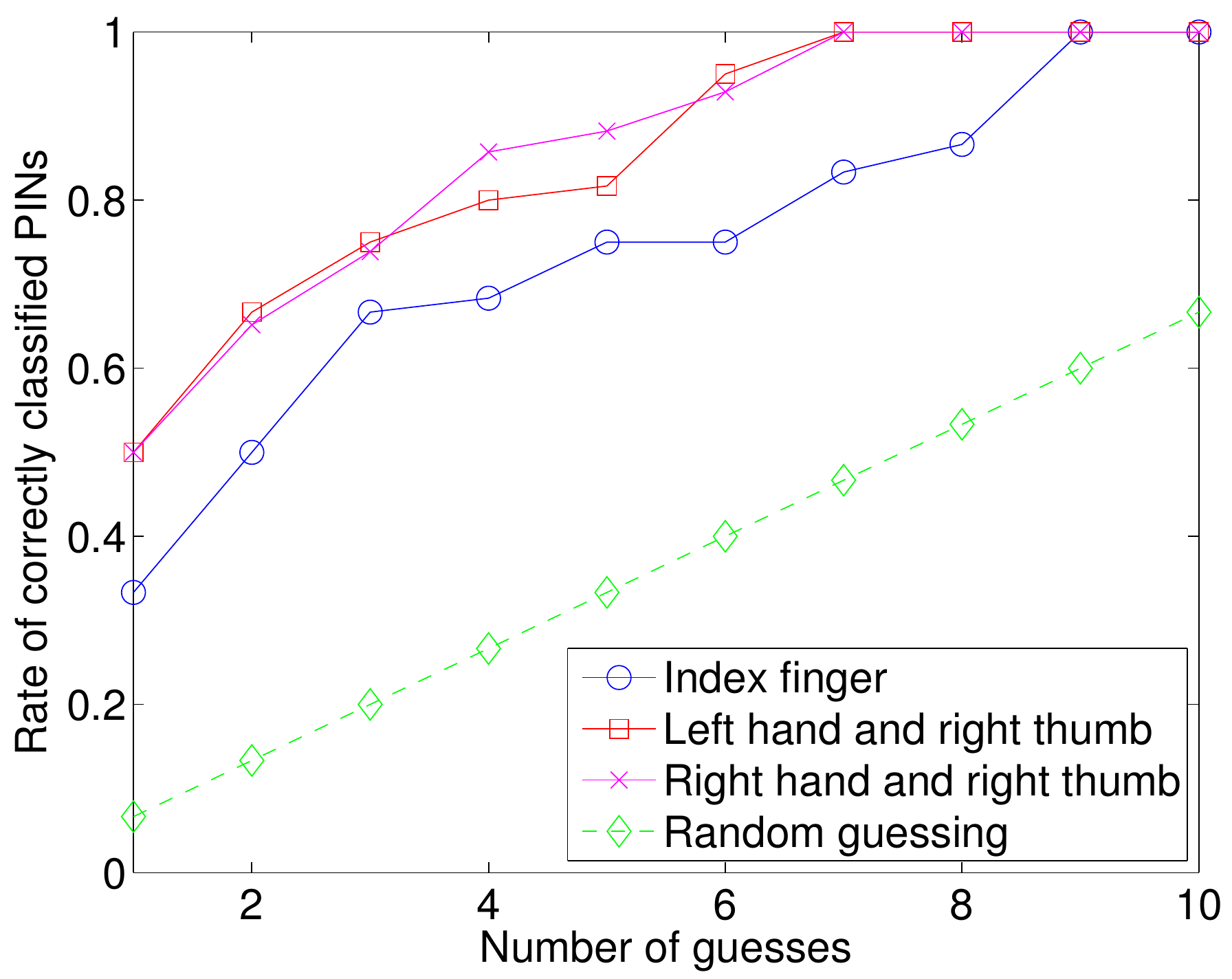}
\caption{Average rate of correctly classified PINs for different input methods.}
\label{fig:classify_correctly_different_input_methods_15_pins}
\end{figure}

\paragraph{Impact of Different Input Methods.}
Since users employ different input methods, we investigate the impact of an input method on the classification rate. To this end, we compare the three major input methods: 
\begin{enumerate}
\item Holding the device in one hand and using the thumb of the same hand to operate it.
\item Holding the device in one hand and operating it with the thumb of the other hand.
\item Holding the device in one hand and using the index finger of the other hand to operate it.
\end{enumerate}

Figure~\ref{fig:classify_correctly_different_input_methods_15_pins} illustrates the average rate of correctly classified PINs for the three different input methods after guessing a specific number of the most probable PINs. The plot is based on a 10-fold cross validation considering a \emph{discriminant analysis} on the LRGBW values. The underlying set of PINs had a cardinality of 15. According to this plot the two input methods involving the thumb, \ie,  left hand and right thumb as well as right hand and right thumb, seem to be more vulnerable to this attack than the one with the index finger. This is due to the fact that the mobile device usually undergoes only minor movements when the index finger is used, because one hand is solely used to hold the mobile device. However, this is not entirely correct because also for the input method involving the left hand and the right thumb one hand is solely used to hold the mobile device.

\setlength{\tabcolsep}{6pt}
\begin{table}[t]
 \centering
\resizebox{2.1in}{!}{
 \begin{tabular}{ll}
 \toprule
\multicolumn{1}{c}{User} & \multicolumn{1}{c}{Input method} \\
 \midrule
User 1 & Left hand and index finger \\
User 2 & Right hand and right thumb \\
User 3 & Left hand and index finger \\
 \bottomrule
 \end{tabular}
}
\caption{Input methods of three users.}
\label{tab:users}
\end{table}

\begin{figure}[!t]
\centering
\includegraphics[width=2.7in]{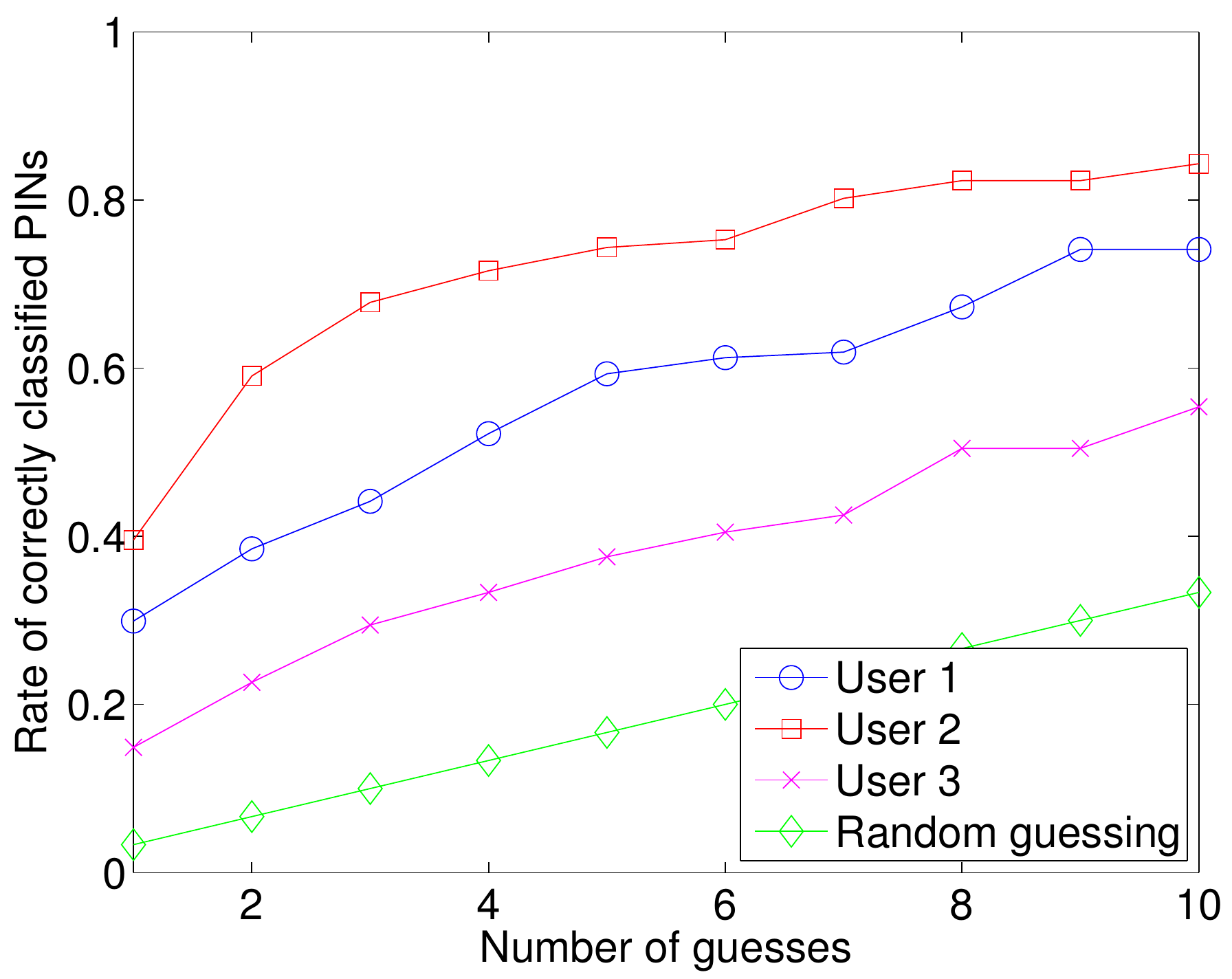}
\caption{Average rate of correctly classified PINs for three specific users on a set of 30 PINs.}
\label{fig:classify_correctly_three_users_30_pins}
\end{figure}

To gain further insight into factors affecting the rate of correctly classified PIN inputs, we compare the provided data of three different users. The corresponding input methods of these users are illustrated in Table~\ref{tab:users}. All three users entered $30\times3$ PINs, \ie, each of the 30 PINs was entered 3 times, within the same room  under the same environmental conditions regarding the ambient light. Figure~\ref{fig:classify_correctly_three_users_30_pins} illustrates the result of the 10-fold cross validation of the three data sets provided by \emph{User 1}, \emph{User 2}, and \emph{User 3}, respectively.  The input method of \emph{User 2} seems to leak the most information. This appears to be due to the fact that she rested her upper arm against her upper body in a relaxed manner and operated the smartphone in a very comfortable way. \emph{User 1} placed her elbows on her knees and also operated the device 
in a very relaxed way. In contrast, \emph{User 3} tightly pressed her upper arm against the upper body and held the device very firmly in her hand while entering the digits with the index finger. Based on the investigations of these three users, we observe that though the input method seems to have an impact on the classification rate, a general statement regarding the security or insecurity of a specific input method is difficult to make. Nevertheless, we claim that the tighter and more firmly one holds the device, the less information is leaked. To put it more generally, the more movements the mobile device undergoes during the operation, the more information is leaked.

\paragraph{Impact of the Sampling Frequency.}
As outlined in Section~\ref{sec:investigation_of_light_sensor}, the ambient-light sensor can be configured to operate with a variety of different sampling frequencies. With a sampling frequency of 750\,Hz the \emph{Samsung Galaxy SIII} provides an immense number of measurement samples per second, far more than is necessary for a successful attack. In fact, most of our attacks were performed with a sampling frequency between 5 and 50\,Hz, though we performed successful attacks with all possible sampling frequencies. 

\begin{figure}[!t]
\centering
\includegraphics[width=2.7in]{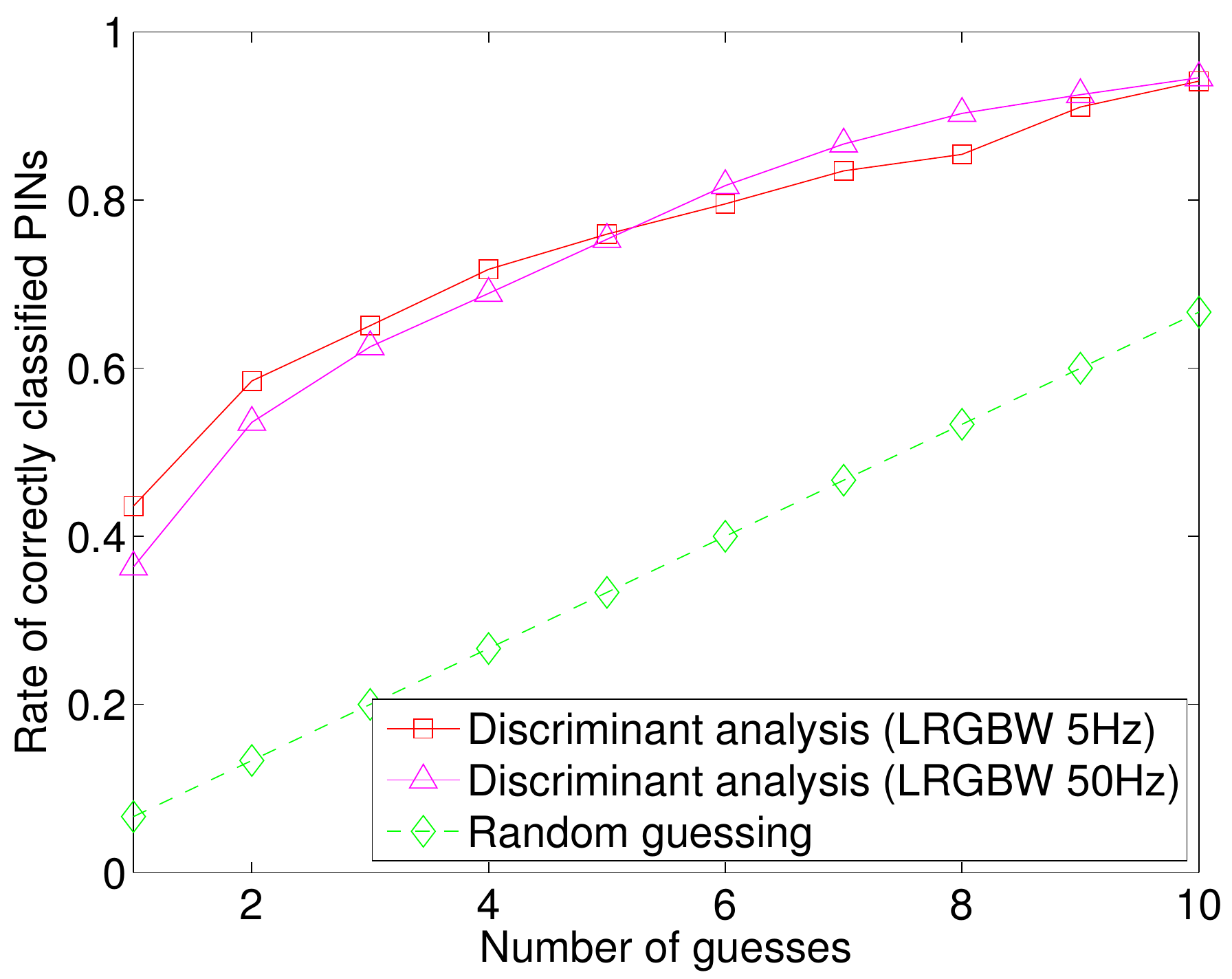}
\caption{Average rate of correctly classified PINs for different sampling rates based on a set of 15 PINs.}
\label{fig:classify_correctly_different_sampling_frequencies_15_pins}
\end{figure}

Figure~\ref{fig:classify_correctly_different_sampling_frequencies_15_pins} indicates that the lowest sampling frequency supported by our device (5\,Hz) is actually enough to perform the presented attack. The plot illustrates that the performance does not even decrease when sampling with the lowest frequency of 5\,Hz. 

Demonstrating that an attack can be performed with a low sampling frequency is more important than demonstrating its success on a high sampling frequency. This is due to the fact that multiple devices are already equipped with such a sensor and upcoming mobile devices are most likely equipped with more powerful RGB(W) sensors, \ie, supporting higher sampling frequencies and resolutions. 

Though we did not observe any problems with the lowest sampling frequency on the \emph{Samsung Galaxy SIII}, we note that even lower sampling frequencies potentially might prevent this attack. This is due to the fact that for sampling frequencies below 5\,Hz too few measurement samples might be gathered if one enters the PIN too fast. 

\section{Limitations}
  \label{sec:limitations}
The presented attack also has some limitations. First, our model does not consider mistyped PINs. If a user deletes an incorrect digit and enters the correct digit afterwards we are not able to infer the correct PIN anymore. Nevertheless, related work does not cover this case neither. Second, we did not evaluate our attack outside under sunny, foggy or cloudy light conditions. However, we evaluated the attack in a room---where the only light source was a window---during different daytimes. Furthermore, the data used to train the machine-learning algorithm must be gathered in the same environment as the actual data that should be classified. Nevertheless, our attack scenario is based on the assumption that people use their smartphone while watching TV, while waiting in a waiting room, or during a train ride, which seems to be a reasonable assumption. Future work, however, might investigate whether a more general model can be established in order to decouple the training phase from the actual attack phase. For instance, a ``calibration'' phase might be used to determine the overall light conditions in the user's environment to speed-up the training phase. Third, due to the fact that the presented attack is based on the ambient-light sensor, it does not work in case the user operates the mobile device in a completely dark environment. Though, a completely dark environment also prevents an attack that exploits the camera~\cite{Simon:2013:PSI:2516760.2516770}.

\section{Analysis of Countermeasures}
  \label{sec:analysis_countermeasures}
In this section, we discuss potential mitigation techniques to prevent the exploitation of sensor information.

\paragraph{UI and API Modifications.} 
Aviv~\etal~\cite{DBLP:conf/acsac/AvivSBS12} argue that an effective security mechanism would be to prevent untrusted applications from accessing motion sensors, at least during the input of sensitive information. However, the crucial question is: \emph{When is an input considered as being sensitive?} Clearly, the input of an authentication PIN or a password represents a sensitive input. But what about the input while writing an e-mail or the data entered in forms on websites? Perhaps the sensors should be disabled as soon as the virtual keyboard is displayed? However, this renders applications that rely on these sensors completely useless as is also stated by Aviv~\etal~\cite{DBLP:conf/acsac/AvivSBS12}. 

Owusu~\etal~\cite{DBLP:conf/wmcsa/OwusuHDPZ12} suggest to vary the keyboard layout for sensitive inputs, which means that buttons are rearranged on the virtual keyboard prior to every sensitive input. The drawback of such countermeasures is the dramatic decrease of usability. While this might be applicable for a PIN pad, it would definitely undermine the usability of the QWERTY keyboard layout.

Limiting the resolution and the sampling frequency of the sensor might be another possible countermeasure. For instance, the ambient-light sensor is currently used to adapt the screen brightness. For such an application a more coarse resolution as well as a lower sampling frequency should be sufficient. We are not aware of any scenario that requires a sampling frequency of 50\,Hz or even 750\,Hz. As we have shown in this paper, even the lowest sampling frequency of 5\,Hz on the \emph{Samsung Galaxy SIII} does not prevent the presented attack. However, reducing the sampling frequency to 1-2\,Hz should suffice for the task of adapting the screen brightness and to hedge the presented attack. Furthermore, since only the OS performs the task of adapting the screen brightness, access to this sensor might be restricted to the OS exclusively. While these countermeasures might be quite effective, they limit the functionality of specific applications, \eg, games that heavily rely on the usage of sensors. 

\setlength{\tabcolsep}{6pt}
\begin{table*}[t]
 \centering
 \resizebox{6.5in}{!}{
  \begin{tabular}{llll}
 \toprule
  & Aviv~\etal~\cite{DBLP:conf/acsac/AvivSBS12} & Simon and Anderson~\cite{Simon:2013:PSI:2516760.2516770} & Ours\\
 \midrule
\textbf{Sensor} & Accelerometer & Camera & Ambient-light sensor \\
\textbf{Permissions} & Internet & Camera, Internet & Internet \\
\textbf{Training} & Independent of user/location & For each user individually & For each user/environment individually \\
\textbf{Drawbacks} & - & LED and shutter sound on non-rooted devices & Does not work in completely dark environments \\
\textbf{Input method} & No constraints & Thumb of holding hand & No constraints \\
\textbf{Accuracy} & 43\% within 5 guesses & 50\% within 5 guesses & 65\% within 5 guesses \\
 \bottomrule
 \end{tabular}
}
\caption{Comparison of related work targeting a set of 50 PINs.}
\label{tab:related_work}
\end{table*}

\paragraph{Rethinking the Permission Model.}
A quite sophisticated countermeasure might be a fine-grained permission system in mobile operating systems. Felt~\etal~\cite{Felt:2012:AP:2372387.2372394} evaluated different permission-granting mechanisms including automatic granting, trusted UIs (cf. Roesner~\etal~\cite{DBLP:conf/sp/RoesnerKMPWC12}), runtime-consent dialogs, and install-time warnings. After considering their model we conclude that an effective countermeasure would be an install-time warning, \ie, to pause the installation process and to explicitly inform or warn the user about the requested permission. 

Specific risks that might arise from permissions must be communicated to the user, as has also been reported by Felt~\etal~\cite{DBLP:conf/soups/FeltHEHCW12}, especially since the manifold permissions confuse many users~\cite{DBLP:conf/soups/FeltHEHCW12,DBLP:conf/fc/KelleyCCJSW12}. However, excessive warnings lose effectiveness and might cause users to ignore these warnings again. In order to overcome this problem Peng~\etal~\cite{DBLP:conf/ccs/PengGSLQPNM12} suggest to rank applications according to their risks rather than using a binary decision for the classification of vulnerable applications. This ranking decision is based on the requested permissions of applications that are known to be malware. Based on such a ranking users might make more deliberate decisions regarding the installation of applications. However, such rankings are only applicable if the motion sensors as well as the ambient-light sensor are considered within the permission system of the OS, which calls for security-specific actions of operating-system developers.

\paragraph{Application Analysis.}
A similar approach might be achieved by extending \emph{AppGuard}~\cite{DBLP:conf/tacas/BackesGHMS13}---an Android application to enforce security policies---to support the detection of possibly unwanted sensor accesses by malicious applications. \emph{AppGuard} could scan applications during the installation and inform the user about sensor accesses that potentially leak sensitive information. Other malware-analysis applications such as static analyzers, \eg, \emph{Stowaway}~\cite{DBLP:conf/ccs/FeltCHSW11} or \emph{AndroidLeaks}~\cite{DBLP:conf/trust/GiblerCEC12}, or applications like \emph{VirusTotal}~\cite{AndroidVirusTotal} 
could also be extended to check  applications for malicious sensor accesses. 

\paragraph{User Behavior.}
Another possible countermeasure might be to enter sensitive data only in environments without any light source and with the index finger or a stylus pen. However, in this case other sensors, \eg, motion sensors, might still be exploitable. So, for really sensitive data, the user might cover the ambient-light sensor as well as the camera, \eg, with her finger, and place the mobile device on a flat surface while providing the input. 

Encouraging users to choose longer PINs and passwords might also increase the security~\cite{Simon:2013:PSI:2516760.2516770}. However, some applications do not even allow PINs with more than 4--5 digits. Again, the drawback of such countermeasures is the decreasing usability.

Last but not least, awareness must be raised amongst users. Applications should not be able to gather information without knowledge of the user and 
users must be encouraged to be wary when installing applications, which is why studies like this one are essential.

\section{Related Work}
 \label{sec:related_work}
Side-channel attacks on input devices in general have been shown to occur both directly~\cite{DBLP:conf/sp/AsonovA04, DBLP:conf/ccs/KuneK10,DBLP:conf/uss/SongWT01} as well as indirectly, \eg,  through oily residues on the touchscreen~\cite{Aviv:2010:SAS:1925004.1925009}, or through reflections of monitors~\cite{DBLP:conf/sp/BackesCDLW09, DBLP:conf/sp/BackesDU08}. More specific investigations of threats represented by sensor-based side channels on mobile devices include, for instance, the work of Cai~\etal~\cite{DBLP:conf/sigcomm/CaiMC09}. They raised the awareness regarding the camera, the microphone, and the GPS signal in modern smartphones. In 2011, Cai and Chen~\cite{Cai:2011:TIK:2028040.2028049} claimed to be the first to show the privacy risk of motion sensors utilized in smartphones. Han~\etal~\cite{DBLP:conf/comsnets/HanONPZ12} used the accelerometer sensor to infer the location of the device owner, even with the location-based services deactivated. In 2012, Owusu~\etal~\cite{DBLP:conf/wmcsa/OwusuHDPZ12} employed the accelerometer sensor to infer passwords entered on touchscreens. Miluzzo~\etal~\cite{DBLP:conf/mobisys/MiluzzoVBC12} and Xu~\etal~\cite{DBLP:conf/wisec/XuBZ12} made use of the accelerometer and motion sensors in general to infer the locations of taps on touchscreens. Recently, attacks have been presented to infer unknown PIN inputs based on a set of learned PINs. For instance, Aviv~\etal~\cite{DBLP:conf/acsac/AvivSBS12} demonstrated such an attack by exploiting the accelerometer and Simon and Anderson~\cite{Simon:2013:PSI:2516760.2516770} demonstrated such an attack by exploiting the camera and the microphone. We briefly compare related attacks to our attack within the following paragraphs. 

\paragraph{Comparison.}
Compared to attacks based on motion sensors, a major advantage of the ambient-light sensor is the power consumption. The method \emph{android.hardware.Sensor.getPower()} returns the power in \emph{mA} used by the corresponding sensor while in use~\cite{AndroidDevelopersSensors}. On our \emph{Samsung Galaxy SIII} this method returns $0.2$\,\emph{mA} for the ambient-light sensor, $0.23$\,\emph{mA} for the accelerometer, and $6.1$\,\emph{mA} for the gyroscope. Hence, the ambient-light sensor consumes a factor of 30 less power than the gyroscope, which means that our attack is less prone to gain the user's attention through battery drainage.

Table~\ref{tab:related_work} provides a comprehensive comparison of the related work that is similar to ours, \ie, attacks targeting a specific set of PINs. All attacks assume to have Internet access in order to transfer the gathered data to a powerful server that performs the machine learning. However, as argued in Section~\ref{sec:attack_scenario}, the \emph{Internet} permission can be gained rather easily without raising the user's suspicion. 

The main drawback of our attack is that users are not allowed to walk around while entering the PINs because the data is only exploitable for one specific environment. Hence, the data cannot be reused for multiple attacks as in case of Aviv~\etal~\cite{Aviv:2010:SAS:1925004.1925009}. Though, their results indicate a rather low success rate of 20\% within 5 guesses when inferring PINs that were entered while walking around. Furthermore, their attack also works in completely dark environments which does not hold for the ambient-light sensor. However, our results indicate a better accuracy of 65\% within 5 guesses when inferring unknown PINs. 

The work of Simon and Anderson~\cite{Simon:2013:PSI:2516760.2516770} additionally requires the \emph{Camera} permission which potentially gains the user's suspicion. In addition, their attack must deal with the problem of audio-visual feedback, \eg, the shutter sound or the LED, while capturing the required data. Compared to their work we do not restrict our study to specific input methods as long as the user holds the device while operating it. Furthermore, they need to transfer image data to the server, which cannot be represented as compact as sensor values.

To summarize this comparison, an attacker trades a higher classification rate for two minor drawbacks when using the ambient-light sensor.

\section{Conclusion and Future Work}
\label{sec:conclusion}
In this paper, we investigated a new type of side channel which is based on the ambient-light sensor employed in today's mobile devices. To the best of our knowledge, we are the first to show that the ambient-light sensor indeed leaks sensitive information about the user's input on the touchscreen. We developed a proof-of-concept application that allows us to infer unknown PINs, when given a set of already known PINs. This application clearly demonstrates that the leaked information represents a viable side channel for compromising the user's privacy and security. Since no specific permission is required to access this sensor, an adversary is able collect sensitive inputs from mobile-device owners without raising any suspicion and, thus, remain unrecognizable. 

Any mobile device equipped with an ambient-light sensor that provides a sufficiently high sampling rate and resolution can be exploited. Our investigation showed that state-of-the-art smartphones---from all major device manufacturers---include an ambient-light sensor and the rapid technical progress in combination with demanding users possibly leads to an increasing resolution and sampling frequency of this sensor. For a list of potential vulnerable devices see Appendix~\ref{sec:devices_potentially_at_risk}.

There are many different dimensions among the information leakage of sensors can be investigated. Examples of factors affecting the applicability and performance of sensor-based side-channel attacks are, for instance, the sensor hardware itself, the screen dimension, the device orientation, the keyboard layout, the user's behavior and his typing style, the different classification algorithms and the employed feature vectors, the actual environment (\eg, indoor and outdoor), etc. 
These examples demonstrate that further research is necessary to evaluate the performance of all available sensors under different settings in order to determine the best sensor for a specific attack scenario. For instance, Cai and Chen~\cite{DBLP:conf/trust/CaiC12} as well as Al-Haiqi~\etal~\cite{al2013keystrokes} performed a comparison of the gyroscope and the accelerometer for multiple users on different devices in different settings to determine the best motion-based sensor. 
It remains an open question whether a combination of motion sensors with the ambient-light sensor might lead to better performances than reported so far. However, the intention of this work was to provide a first feasibility study. By comparing three classification algorithms, different feature vectors, different input methods, different environments, and the impact of different sampling frequencies, we showed that the ambient-light sensor provides a viable side channel.

Related work on the leakage of motion sensors claimed that access to these sensors must be limited with a fine-grained permission system. As shown in this work, access to the ambient-light sensor must also be protected through such a permission system and, thus, operating-system developers need to deal with this problem. Probably even more important is the fact that users need to be aware of such threats and be wary when installing applications that require permissions to access sensors. This is actually why studies like this one are essential.

\setlength{\tabcolsep}{6pt}
\begin{table*}[b]
 \centering
\resizebox{4.7in}{!}{
  \begin{tabular}{llrr}
 \toprule
 Device & Operating system & Sample rate [Hz] & Resolution [lux] \\
 \midrule
Google Nexus S         & Android 2.3.6     & $\sim 20$  & \{10, 160, 320, 640\} \\
Google Nexus S         & Cyanogenmod 4.0.4 & $\sim 140$ & 1 \\
HTC Nexus One          & Android 2.3.6     & $\sim 1$   & 1 \\
LG Optimus G           & Android 4.1.2     & $\sim 7$   & 1 \\
LG Optimus G Pro       & Android 4.1.2     & $\sim 7$   & 1 \\
Samsung Galaxy SII     & Android 2.3.4     & $\sim 10$  & \{10, 100, 1000\} \\ 
Samsung Galaxy SIII    & Android 4.3       & $\sim 750$ & 1 \\
Samsung Galaxy S4 Mini & Android 4.2.2     & $\sim 100$ & 1 \\  
Samsung Galaxy Note II & Android 4.3       & $\sim 100$ & 1 \\
 \bottomrule
 \end{tabular}
}
\caption{Mobile devices and their corresponding operating system as well as the observed sampling rate and the resolution of the ambient-light sensor.}
\label{tab:devices_at_risk}
\end{table*}

{\footnotesize \bibliographystyle{acm}
\bibliography{template}}

\begin{appendix}

\section{Mobile Devices Potentially at Risk}
  \label{sec:devices_potentially_at_risk}
Table~\ref{tab:devices_at_risk} lists Android-based mobile devices as well as the empirically determined sampling frequencies and resolutions of the light sensor.
For the \emph{Google Nexus S} running Android 2.3.6 and the \emph{Samsung Galaxy SII} running Android 2.3.4 we state the observed \emph{lux} values instead of the actual resolutions.

Except the \emph{HTC Nexus One}, all investigated devices support a sampling rate that is high enough for the presented attack. For two devices, \ie, the \emph{Google Nexus S} running Android 2.3.6 and the \emph{Samsung Galaxy SII} running Android 4.3, the limiting factor seems to be the resolution. The limited resolution in turn seems to be a result of the employed operating system rather than the hardware itself. For instance, the \emph{Google Nexus S} running Cyanogenmod 4.0.4 provides a more fine-grained resolution than the \emph{Google Nexus S} running Android 2.3.6. 
This also complies with our observation that reading the sensor values directly from the virtual file system, \eg, \emph{/sys/devices/virtual/lightsensor/switch\_cmd} on the \emph{Samsung Galaxy SII}, provides a better resolution than using the official Android Sensor API. 

Since this table contains only devices we had at hand, we expect far more devices to be vulnerable to attacks based on the ambient-light sensor. 

\end{appendix}

\end{document}